\documentclass[12pt]{iopart}

\usepackage{iopams}  
\usepackage{graphicx}
\usepackage{srctex}
\usepackage[dvipsnames]{xcolor}

\begin{document}

\title[]{Synchronization and collective motion of globally coupled Brownian particles}

\author{Francisco J.\ Sevilla}
\ead{fjsevilla@fisica.unam.mx}
\address{Instituto de F\'isica, Universidad Nacional Aut\'onoma de M\'exico,\\
Apdo. Postal 20-364, 01000, M\'exico D.F., M\'exico}

\author{Victor Dossetti}
\ead{dossetti@ifuap.buap.mx}
\address{ Instituto de F\'isica, Benem\'erita Universidad Aut\'onoma de Puebla,\\
Apdo.\ Postal J-48, Puebla, Pue. 72570, M\'exico and\\ Consortium of the Americas for Interdisciplinary Science,\\ University of New Mexico, Albuquerque, NM 87131, USA}

\author{Alexandro Heiblum-Robles}
\ead{heiblum@fisica.unam.mx}
\address{Instituto de F\'isica, Universidad Nacional Aut\'onoma de M\'exico,\\
Apdo. Postal 20-364, 01000, M\'exico D.F., M\'exico}

\begin{abstract}
In this work, we study a system of passive Brownian (non-self-propelled) particles in two dimensions, interacting only through a social-like force (velocity alignment in this case) that resembles Kuramoto's coupling among phase oscillators. We show that the kinematical stationary states of the system go from a phase in thermal equilibrium with no net flux of particles, to far-from-equilibrium phases exhibiting collective motion by increasing the coupling among particles. The mechanism that leads to the instability of the equilibrium phase relies on the competition between two time scales, namely, the mean collision time of the Brownian particles in a thermal bath and the time it takes for a particle to orient its direction of motion along the direction of motion of the group. 
\end{abstract}

\maketitle

\section{Introduction}

The study of the emergence of collective phenomena in systems far from equilibrium has been developed during the last three decades along different paths. One of these, corresponds to globally interacting entities which develop in time a variety of synchronized collective behaviors such as the synchronized flashing in swarms of some species of fireflies \cite{Buck1988} or the chorusing behavior observed in groups of crickets \cite{WalkerScience1969, SismondoScience1990}. Another example corresponds to the appearance of collective motion in systems of active particles interacting through short-ranged velocity-aligning forces as is thought to occur, in a simplified manner, in flocks of birds \cite{Bajec2009} or schools of fish \cite{Parrihs2002} among others. Though motion is not relevant in the first case, it is evidently in the second one.  

At a first glance, these two mechanism for collective phenomena may seem to be unrelated, however, in both cases, individual behavior is changed in favor of a collective one that emerges from the interactions among individuals. Thus, while synchronization corresponds to the emergence of a collective rhythm among the many ones that characterize each element in the system, collective motion results from the appearance of a global state in which particles move towards a collectively determined direction. 

On the other hand, the collective motion exhibited in many real systems makes us to think of the whole system (some times formed by thousands of individuals \cite{Ballerini2008}) as a single self-propelled entity that behaves coherently. In such a case, one may conceive that the mechanism for the emergence of such behavior lies on the effects of a synchronizing force among the elements of the system. Thus, a plausible bridge between the synchronizing behavior and collective motion, if any, must be unveiled.   

The synchronized behavior exhibited by a collection of globally interacting phase-oscillators is described by the paradigmatic model of Kuramoto \cite{Kuramoto1975, Kuramoto1984, StrogatzPhysicaD2000, AcebronRMP2005}, where the system suffers a dynamic transition as the intensity of the coupling among the oscillators is increased, going from phases where the elements oscillate independently with their own pace, to phases where the elements oscillate synchronously with the same collectively-developed frequency. As such, collective synchronization of many realistic systems has been understood in the context of this exhaustively studied model and extensions of it \cite{AcebronRMP2005, AcebronPRL1998, HongJPhysA1999, AcebronPRE2000}, and one would expect it could be extended to other phenomena such as collective motion \cite{HaNonlinearity2010}. 

In general, the onset of collective motion in systems consisting of \emph{active} or \emph{self-propelled} particles is exhibited by the spontaneous emergence of ordered states in which particles move roughly about the same instantaneous direction. For collective motion, the counterpart of the model of Kuramoto corresponds to the exhaustively studied model by Vicsek \etal \cite{VicsekPRL95}, which exhibits the emergence of a collective behavior in a system of particles that move with constant speed, with dynamics driven by a set of automata-like rules. The problem of how the information is transmitted through the system when its elements interact via short-ranged forces has been addressed in many different studies; see, for example, \cite{HelbingRMP2001, VanniPRl2011}.

Self-propulsion, in its more general meaning \cite[and references therein]{RomanczukEPJ2012}, refers to out-of-equilibrium dynamics in which the energy of the surroundings is turned into particle motion. This mechanism has been thought an essential ingredient for the existence of phases with coherent motion. Indeed, it is large the number of studies on flocking behavior where the self-propulsive character of the agents is kept as an important aspect of the analysis (see for instance \cite{GrossmannNJP2012}), sometimes making the modeling of this kind of behavior complex, with the use of sophisticated nonlinear friction terms \cite{ErdmannPRE2002,ErdmannPRE2005}. This bias may be justified on the basis that the concept of self-propelled particles captures the natural ability (as seen in many biological systems) for the agents to develop motion by themselves \cite{GrossmannNJP2012, SchweitzerBook}. Additionally, it has also been thought to be an important ingredient for pattern formation in models of collective motion \cite{ShimoyamaPRL2000, LevinePRE, EbelingComplexity2003, ToumaPRE2010, DossettiPRE2009}. Thus, a natural question would be, up to what extent self-propulsion is a necessary feature for a system to develop collective motion?

On the other hand, it seems natural to make an attempt to build a theoretical framework bridging between the commonly understood concepts of \emph{synchronization} and \emph{collective motion}. For this,  we must first take into account the subtle but important difference between both: in synchronization, a collective frequency emerges from distinct ones that characterize each one of the \emph{globally} interacting elements, while collective motion emerges from identical self-propelled particles that interact through \emph{short-ranged} \textquotedblleft social forces\textquotedblright. Though, not in its more general scope, such a bridge has been addressed before in \cite{ChepizhkoPhysA2010} by connecting the Kuramoto model of synchronization with the Vicsek \etal model of collective motion. 

In addition, the out-of-equilibrium nature of the steady-states (ordered and disordered) reached in systems of \emph{active} particles, which perform out-of-equilibrium dynamics, are of great interest in modern statistical mechanics \cite{ChaudhuriPRE2014}. In contrast, here we put forward the study of the out-of-equilibrium steady-states in systems of \emph{passive} particles that emerge due to the sole effects of social interactions, as the aligning force considered in this work. The transition to these states, marks the breakdown of the fluctuation-dissipation-relation's validity that, in general, corresponds to the emergence of net non-zero probability currents in configuration space.

In the present paper, we focus mainly on three aspects: (i) To make clear that self-propulsion is an unnecessary non-equilibrium feature of the agents to exhibit phases of collective motion. We make explicit this by considering only the underlying dynamics of passive standard Brownian motion in a system of globally interacting particles. In other words, the inertial effects of Brownian motion in addition to velocity-alignment interactions suffice to exhibit collective motion in a system of passive Brownian particles. This conclusion does not rely on  the mean field nature of the interaction as will be shown elsewhere, in a subsequent analysis. In addition, this result suggests the theoretical possibility for engineering systems of \emph{passive} particles to display collective motion. (ii) To show that the model of velocity-alignment among particles introduced in this work, which separates the dynamics of turning from the dynamics of propulsion (Brownian in the present case), allows us to discuss the precise relationship of the appearance of collective motion with the onset of collective synchronization. Particularly, it allows us to make a direct connection with the model of Kuramoto, in contrast with other studies \cite{GrossmannNJP2012, RatushnayaPhysA2007}, where such relation is just implied. (iii) To show that synchronization implies collective motion, in other words, considering globally coupled identical movers is sufficient to exhibit such collective behavior. 

Though the emergence of collective motion is indeed expected in the mean-field situation proposed here (global coupling), our model allows us to analytically study the transition from steady-states close to equilibrium to steady-states out of equilibrium, by tuning the velocity-alignment among particles alone. Indeed, the transition from stationary \emph{asynchronous} states to stationary \emph{synchronous} ones corresponds, in our model, to the transition from close-to-equilibrium stationary phases to out-of-equilibrium ones, characterized by a particle current and, therefore, the production of entropy. 

We must mention that there has been a relatively recent re-growth of interest in systems with long-range interactions in which, contrasting with systems that consider short-range interactions, unusual effects may arise \cite{CampaPhysRep2009}. An example is the so-called Brownian mean-field model \cite{ChavanisEPJB2014} that can be interpreted as a one-dimensional model of \emph{flocking} behavior with long-range Kuramoto-like interactions.

We address all these aspects by formulating a model based on Langevin equations, for globally interacting Brownian particles for which the interacting mechanism avoids the self-driving characteristic. In section \ref{sec2} we introduce our model in terms of Langevin-like equations and the nature of the velocity-alignment force is discussed. In section \ref{sec4} we analyze the stability of the equilibrium phase, against interactions, by performing a stability analysis of a nonlinear Fokker-Planck equation for the probability density, $P(\boldsymbol{v,t})$, for finding a particle moving with velocity $\boldsymbol{v}$ at time $t$. We finally summarize our conclusion in section \ref{conclusions}.

\section{\label{sec2} Model}
We consider $N$ two-dimensional Brownian particles in the underdamped limit, that interact among themselves through a velocity-aligning mechanism that incorporates a finite aligning rate without affecting the magnitude of the velocity of the particles. Clearly, this contrasts to other studies in continuous time, in that our approach is intended to disentangle the effects of active motion (generally included by non-linear friction forces that drive the particles to move with a speed around a constant value \cite{GrossmannNJP2012} or, in the overdamped limit, to move with constant speed \cite{RatushnayaPhysA2007, FarrellPRL2012}), from the effects of velocity alignment. 

The velocity-alignment interaction is of particular interest since it involves the dynamical attraction of the single particle direction of motion, to one collectively determined by the coupling with the rest of the elements in the system, resembling the synchronizing force in Kuramoto's model of globally-interacting phase oscillators. In this way, two-dimensional systems seem to be important for this matter since a direct connection with Kuramoto's model can be established as shown here. 

\begin{figure}
\includegraphics[width=0.6\textwidth,clip]{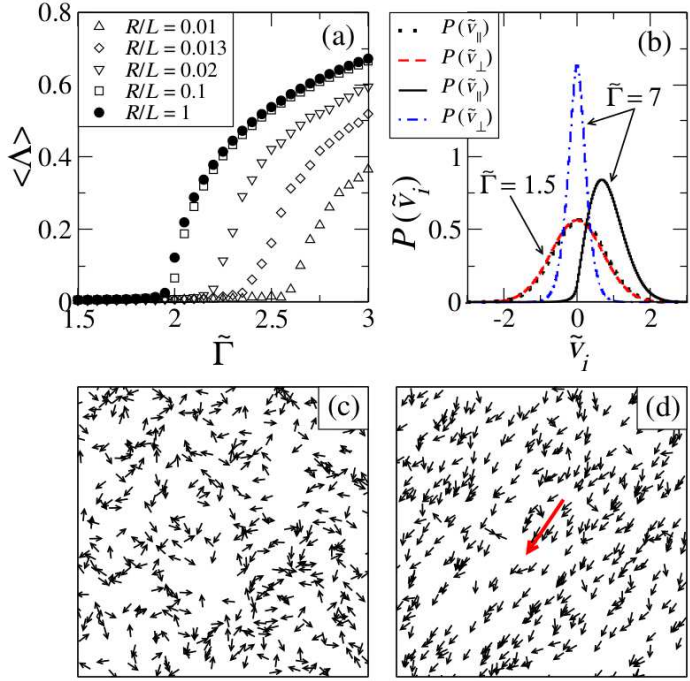}
\caption{(Color online) (a) Stationary order parameter $\langle \Lambda \rangle$ vs the coupling constant $\tilde{\Gamma}$ for different ratios of the interaction radius $R$ and the system size $L$ with $\tilde{\rho}=10$ and $N=10^5$. The system is driven from the local $(R/L<1)$ to the global $(R/L=1)$ coupling regimes as $R/L$ increases. The rest of the plots consider only the global coupling regime. (b) Stationary probability distribution functions of the individual velocity of the particles, $\tilde{\boldsymbol{v}}_i$, projected along the direction of the mean velocity of the group, $\tilde{v}_{\|}$, and in the transverse direction, $\tilde{v}_{\bot}$, for a subcritical $(\tilde{\Gamma}=1.5)$ and a supercritical $(\tilde{\Gamma}=7)$ cases. Notice the asymmetry of $P(\tilde{v}_{\|})$ for the case $\Gamma=7$ which is the only one not centered around zero. In (c) and (d), snapshots of a quarter of the total space for the configuration of particles for subcritical (c) and supercritical (d) cases are shown, corresponding to those presented in (b). The big arrow in (d) depicts the mean direction of motion of the group. The small arrows represent the velocity vectors $\tilde{\boldsymbol{v}}_{i}$ of  the particles. In (b), (c) and (d) $\tilde{\rho}=1$ and $\tilde{L}=40$.}
\label{fig1}
\end{figure}

In the interactionless limit, we assume the dynamics of the Brownian particles to be constrained by the fluctuation-dissipation relation (FDR) as in the standard description of Brownian motion, therefore, the stationary state distribution of the single particle velocities corresponds to that of equilibrium with the highest rotational symmetry exhibited by the circularly symmetric distribution of velocities that correspond to the Gaussian one of Maxwell and Boltzmann. When the interaction among agents is turned on, it is plausible to expect the FDR not to hold and, therefore, there is no guarantee of acquiring a new stationary state of equilibrium. Indeed, aligning behaviors may break the rotational symmetry that characterizes the disordered phases if aligning time-scales are smaller than those related to the FDR. This provides a mechanism for the emergence of synchronized-like behavior of the entire system [Fig.\ \ref{fig1}(d)]. In this case, if initial conditions are compatible with the disordered states, the dynamics exerted by the ordering force would take the system to a state with less rotational symmetry, implying the emergence of a far-from-equilibrium state characterized by a net particle-current. By performing a stability analysis of a nonlinear Kramer-Fokker-Planck equation in the limit of global coupling, we show that the equilibrium state is stable against the aligning force up to a critical value $\Gamma_{c},$ at which, a phase transition takes place.

The model under study is described in terms of generic stochastic differential equations for underdamped Brownian particles of mass $m$, restricted to move within a box of linear size $L$ with periodic boundary conditions, namely
\numparts
\begin{eqnarray}
\frac{d\boldsymbol{v}_{i}}{dt}= \boldsymbol{F}_{i}-\gamma \boldsymbol{v}_{i}+\boldsymbol{\xi}_{i};\label{vel_eq}\\
\frac{d\boldsymbol{x}_{i}}{dt}=\boldsymbol{v}_{i}\label{pos_eq},
\end{eqnarray}
\endnumparts
where $\boldsymbol{F}_{i}$ is the velocity-aligning field that depends only on the velocities of the particles. The last two terms on the right hand side of  (\ref{vel_eq}) correspond to the linear-dissipative force per unit mass and the fluctuating one that appear in the Langevin's description of Brownian motion. Notice that each term in the right hand side of \eref{vel_eq} has been rescaled with the mass of the particle, $m$, thus each term has units of force per unit mass. The components of the vector $\boldsymbol{\xi}_{i}$ are nothing else than Gaussian white noise with vanishing mean and  autocorrelation function $\langle\xi_{i\mu}(t)\xi_{j\nu}(s)\rangle=\delta_{i,j}\delta_{\mu,\nu}2k_{B}T\gamma\delta(t-s)/m$, where $\xi_{i,\eta}$ is the $\eta$-th Cartesian component of $\boldsymbol{\xi}_{i}$, $k_{B}$ is the Boltzmann constant, $T$ the bath's temperature and $\delta_{u,w}$ and $\delta(\tau)$ are the Kronecker delta and the Dirac delta function, respectively. This kind of fluctuations are called passive \cite{RomanczukPRL2011}, as they do not take part in any effect of particle-propulsion. In the absence of interactions, the particle dynamics is driven by thermal fluctuations as occurs in equilibrium phenomena.

It is the underdamped nature of \eref{vel_eq} and \eref{pos_eq} what allows us to consider only the simple Brownian dynamics exerted by thermal fluctuations, without having to take into consideration any self-propulsion mechanism (usually constant speeds) as required in many other models that regard the overdamped limit \cite{FarrellPRL2012, ChepizhkoPRL2013}. On the other hand, the effects of thermal fluctuations in the interactionless limit are well known, and lead (in unbounded space) to a mean squared displacement $\langle[\boldsymbol{x}(t)-\langle\boldsymbol{x}(t)\rangle]^{2}\rangle=(4D/\gamma)\left[\gamma t-(1-e^{-\gamma t})\right]$, where the diffusion constant is given by $D=k_{B}T/m\gamma.$ 
 
The alignment behavior among particles is taken into account by 
\begin{equation}\label{Aligning}
\boldsymbol{F}_{i}=\Gamma(v_{i})\left[\boldsymbol{f}-\hat{\boldsymbol{v}}_{i}\left(\boldsymbol{f}\cdot\hat{\boldsymbol{v}}_{i}\right)\right],
\end{equation}
which forces the alignment of the vector $\hat{\boldsymbol{v}}_{i}$ to the vector $\boldsymbol{f}$ at the speed-dependent rate per unit mass and unit velocity $\Gamma(v_{i})$. The explicit dependance of $\Gamma(v)$ on $v$ may vary from case to case. For instance, one possibility is to chose $\Gamma(v)\sim v^{-\alpha}$ with $\alpha>0$ to model the effects of ``aligning inertia'', or the effect that speedy particles does no have enough time to align with sluggish ones. \eref{Aligning} corresponds to the two-dimensional form of $\Gamma(v_{i})\left[\hat{\boldsymbol{v}}_{i}\times(\boldsymbol{f}\times\hat{\boldsymbol{v}}_{i})\right]$, with $\hat{\boldsymbol{v}}_{i}=(\cos\theta_{i},\, \sin\theta_{i}),$ being the unitary vector in the direction of $\boldsymbol{v}_{i}$ and $v_{i}=\vert\boldsymbol{v}_{i}\vert$, while
\begin{equation}
 \boldsymbol{f}=\frac{1}{N}\sum_{j=1}^{N}\hat{\boldsymbol{v}}_{j},
\end{equation}
corresponds to the ins\-tan\-ta\-neous average direction of motion of the group. As defined, $\boldsymbol{f}$ corresponds also to the instantaneous order-parameter which can be rewritten in a suitable manner in the complex plane as 
\begin{equation}\label{ComplexOP}
\Lambda(t)\, e^{i\psi(t)}=\frac{1}{N}\sum_{j=1}^{N}\, e^{i\theta_{j}(t)},
\end{equation} 
with $\theta_{j}(t)$ being the instantaneous angle between the velocity vector $\boldsymbol{v}_{j}$ and the horizontal axis. The magnitude of $\boldsymbol{f}$,  $\Lambda(t)$, ranges between zero and one and measures the degree of \textquotedblleft collectivity\textquotedblright in the system, while $\psi(t)$ denotes the instantaneous direction of motion of the whole group. This quantities corresponds in an exact way to the order parameter used in the well known \hbox{Kuramoto} model of collective \emph{syn\-chro\-ni\-za\-tion} \cite{AcebronRMP2005}.

Though \eref{vel_eq} and \eref{pos_eq} are generic for the description of active particles \cite{RomanczukEPJ2012}, it is the explicit form of \eref{Aligning} which make them particularly appealing, since the interaction \eref{Aligning} does not change the speed of the particles, therefore it avoids any self-propulsion effect. It is straightforward to check the lack of propulsion in the alignment interaction given in \eref{Aligning} by computing $\boldsymbol{F}_{i}\cdot\hat{\boldsymbol{v}}_{i}=0$. Additionally,  
$\boldsymbol{F}_{i}\cdot\hat{\boldsymbol{\theta}}_{i}=\Gamma(v_{i})\boldsymbol{f}\cdot\hat{\boldsymbol{\theta}}_{i}$, with $\hat{\boldsymbol{\theta}}_{i}$ being the clockwise orthogonal unitary vector to $\hat{\boldsymbol{v}}_{i}$. With these considerations, \eref{vel_eq} and \eref{pos_eq} can be written as 
\numparts
\begin{eqnarray}
\frac{d\boldsymbol{x}_{i}}{dt}&=v_{i}\hat{\boldsymbol{v}}_{i},\label{xpolar} \\
\frac{dv_{i}}{dt}&=-\gamma v_{i}+\frac{k_{B}T\gamma}{m}\frac{1}{v_{i}}+\xi_{v_{i}}, \label{speed}\\
\frac{d\theta_{i}}{dt}&=\frac{\Gamma(v_{i})}{v_{i}}\, \frac{1}{N}\sum_{j=1}^{N}\sin[\theta_{j}-\theta_{i}]+\frac{1}{v_{i}}\, \xi_{\theta_{i}}, \label{direction}
\end{eqnarray}
\endnumparts
where $\xi_{v}$ and $\xi_{\theta}$ are independent Gaussian white noises with autocorrelation function $2k_{B}T\gamma\delta(t-s)/m$, respectively. The second term in expression \eref{speed} comes from the Ito's calculus when performing the change to polar coordinates \cite{GardinerBook}. Notice that extensions of the Kuramoto model that consider inertial effects lead to simplified versions of \eref{speed} and \eref{direction} \cite{AcebronPRL1998,AcebronPhysD2000,GuptaJStatMech2014}.  

We point out that, written in that way, \eref{speed} and \eref{direction} can be set to consider self-propulsion and/or active fluctuations \cite{RomanczukPRL2011, RomanczukEPJ2012} instead of the passive ones that originate in the thermal bath. If necessary, self-propulsion can be incorporated in a simple way by just replacing the constant friction coefficient $\gamma$ by the nonlinear one $\gamma(v_{i})$. The term $-\gamma(v_{i})v_{i}$ would be able to keep the particle speed around a fixed value $v_{0}$ \cite{RomanczukEPJ2012}. Additionally, the noise terms can be replaced by nonthermal independent stochastic processes corresponding to active fluctuations.
As a simple example, in the overdamped limit, the speed of the particles can be set to $v_{0}$ and \eref{xpolar}, \eref{speed} and \eref{direction} reduce to 
\numparts
\begin{eqnarray}
\frac{d\boldsymbol{x}_{i}}{dt}&=v_{0}\hat{\boldsymbol{v}}_{i},\label{xpolar2} \\
\frac{d\theta_{i}}{dt}&=\frac{\Gamma(v_{i})}{v_{i}}\, \frac{1}{N}\sum_{j=1}^{N}\sin[\theta_{j}-\theta_{i}]+\frac{1}{v_{i}}\, \tilde{\xi}_{\theta_{i}}, \label{direction2}
\end{eqnarray}
\endnumparts
where $\tilde{\xi}_{\theta_{i}}$ are non-thermal stochastic processes. The last equations, with coupling functions that also depend on the spatial coordinates to consider short-range interactions, have been the starting point of some studies of collective motion of active particles \cite{FarrellPRL2012, PeruaniEJPST2008}. We want to point out that the use of more general coupling functions---other than the sinusoidal one that appears in \eref{direction2}---can lead to different non-equilibrium properties of systems consisting of interacting self-propelled particles, a case that will be treated elsewhere. The diffusion properties of noninteracting active particles, described by \eref{xpolar2} and \eref{direction2} with $\Gamma(v)=0$, have been recently studied in \cite{SevillaPRE2014} and differ from the ones close-to-equilibrium considered here.

In addition, notice that by choosing $\Gamma(v)=\Gamma_{1} v$ with $\Gamma_{1}$ a positive constant, equations \eref{speed} and \eref{direction} get naturally decoupled, and leads---in the overdamped limit---to the noisy Kuramoto model of phase oscillators with vanishing natural frequencies \cite{AcebronRMP2005, GuptaJStatMech2014}. Instead, in this work we consider the simple case of a speed independent $\Gamma(v)=\Gamma$. 

We choose as time, speed and length scales, the quantities: $\tau_{0}=\gamma^{-1}$, $v_{0}=(2k_{B}T/m)^{1/2}$ and $r_{0}=v_{0}\tau_{0},$ respectively. In this way, the number of parameters in our model is reduced to one, namely, the dimensionless alignment-coupling constant $\tilde{\Gamma}=\Gamma(\tilde{v})\, \tau_{0}/v_{0}$ with $\tilde{v}=v/v_{0}.$ On the other hand, if short-range interactions were to be  considered, two additional parameters appear, namely, the particle density $\tilde{\rho} = N/\tilde{L}^{2}$ and the interaction range $R/L$ with $\tilde{L}=L/r_{0}$.

Our interest lies on the long-time average of $\Lambda(t)$ in the stationary state, denoted here with $\langle\Lambda\rangle$ and given by
\begin{equation}
 \langle\Lambda\rangle=\lim_{T\rightarrow\infty}\frac{1}{T}\int_{0}^{T}\Lambda(t)\, dt.
 \label{sslambda}
\end{equation}
As for the Kuramoto model, this quantity serves here as an order parameter that signals (when  $\langle\Lambda\rangle > 0$) the breaking of the rotational symmetry of the equilibrium phases. Figure \ref{fig1}(a) shows plots of the stationary order parameter $\langle\Lambda\rangle$ as a function of $\tilde{\Gamma}$, computed from numerical simulations. The curves with different symbols corresponds to different values of the interaction range $R/L,$ that goes from small values to the global coupling value $R/L=1$, which corresponds to the case we are interested in this paper. It can be noticed that the system undergoes a phase transition at a critical value, $\tilde{\Gamma}_{c},$ that decreases monotonically  with the range of the interaction $R$, approaching the value 2 as the interaction becomes global. We point out that for the globally coupled case of the Vicsek and Vicsek-related models, the disordered phase (with an order parameter equal to zero) is achieved only with maximal noise in the thermodynamic limit \cite{DossettiPRE2009}, (equivalent to the infinite temperature limit of our model); see, for example, Fig.\ 11 in the reference provided. This fact contrasts with the model presented here, where the disordered or equilibrium phase is still stable for finite values of the coupling constant as our results show.

Snapshots of the stationary phases for globally interacting particles are shown in figures \ref{fig1}(c) and \ref{fig1}(d), for the equilibrium phase with $\tilde{\Gamma}=1.5$, and for the ordered phase showing collective motion with $\tilde{\Gamma}=7$, respectively. The corresponding stationary probability distribution functions of the individual velocity of the particles, $\tilde{\boldsymbol{v}}_i$, projected along the direction of the mean velocity of the group, $\tilde{v}_{\|}$, and in the transverse direction, $\tilde{v}_{\bot}$, are shown in figure \ref{fig1}(b), as labeled in the figure. Notice the asymmetry of $P(\tilde{v}_{\|})$ for the case $\Gamma=7$ which is the only one not centered around zero; here the system presents a non-zero particle-current. Details regarding the numerical integration of the equations of motion are given at the end of section \ref{sec4}.

In the following section we perform a stability analysis to show that the equilibrium distribution is stable in the subcritical regime. This implies a breakdown of the fluctuation-dissipation relation at the critical point $\tilde{\Gamma}_{c}\simeq2$. From our numerical results we find that the Maxwellian distribution of veloci\-ties still holds for all stationary states with $\tilde{\Gamma}<\tilde{\Gamma}_{c}$ as shown for $\tilde{\Gamma}=1.5$ in figure \ref{fig1}(b). Moreover, the single-particle velocity auto-correlation function decays exponentially (not shown in the figure) with the same time scale $\gamma$, just as it occurs in the interactionless limit, implying that the fluctuation-dissipation theorem still holds and can be validly applied. These facts ensure that the system can reach equilibrium in velocity space. In addition, the motion of the particles in this regime is diffusive, with the same diffusion constant $D$ as in the interactionless case, as implied by the same arguments.

\section{\label{sec4} Stability Analysis}

By using the rotational invariance of expression \eref{ComplexOP}, it can be written as
\begin{equation}
\Lambda(t)=\int_{0}^{2\pi}d\theta\, e^{i\theta}\left[\frac{1}{N}\sum_{j=1}^{N}\, \delta\left(\theta-\theta_{j}(t)\right)\right].
\end{equation} 
The term within brackets gives precisely the fraction of particles that move along the $\theta$ direction which in the thermodynamic limit ($N\rightarrow\infty$ and $L\rightarrow\infty$ such that $\rho=N/L^{2}$ is kept constant), can be identified with the probability density of finding a particle moving along the direction $\theta$ at time $t$, $P(\theta,t)$. Thus,
\begin{equation}\label{OP}
\Lambda(t)=\int_{0}^{2\pi}d\theta\, e^{i\theta}P(\theta,t).
\end{equation} 

Since the velocity alignment is global, positions and velocities can be trivially decoupled allowing us to reduce our analysis to a mean-field theory for the single-particle probability density function $P(\boldsymbol{v},t)$. A formal derivation of a Fokker-Planck equation for the single particle probability distribution $P(\boldsymbol{x},\boldsymbol{v},t),$ follows standard procedure \cite{IhlePRE2011,GuptaJStatMech2014}, namely, one starts by deriving the Bogoliubov-Born-Green-Kirkwood-Yvon hierarchy equations for the $N$ interacting particles, and under the assumption of Boltzmann's molecular chaos, we get
\begin{equation}\label{NonLinearKFP1}
\fl\frac{\partial}{\partial t}P(\boldsymbol{v},t)+\nabla_{\boldsymbol{v}}\cdot\left[\Gamma(v)\,\left(\boldsymbol{f}-\hat{\boldsymbol{v}}\left(\boldsymbol{f}\cdot\hat{\boldsymbol{v}}\right)\right)P(\boldsymbol{v},t)\right]
=\nabla_{\boldsymbol{v}}\cdot\left[\gamma\boldsymbol{v}+\frac{\gamma k_{B}T}{m}\nabla_{\boldsymbol{v}}\right]P(\boldsymbol{v},t),
\end{equation}
where $\boldsymbol{f}$ is computed self-consistently through  
\begin{equation}\label{fofP}
\boldsymbol{f}=\int d^{2}\boldsymbol{v}\, \hat{\boldsymbol{v}} P(\boldsymbol{v},t).
\end{equation}
It is precisely this relation the one that leads to the nonlinear character of equation \eref{NonLinearKFP1}. On the other hand, the probability density that appears in expression \eref{OP} is related with the one in \eref{fofP} through: $P(\theta,t)=\int_{0}^{\infty}dv\,v P(v,\theta,t),$ where $ P(v,\theta,t)$ is obtained from $P(\boldsymbol{v},t)$ by changing to polar coordinates. 

In the absence of velocity alignment, equation \eref{NonLinearKFP1} corresponds to the standard linear Fokker-Planck equation, describing the dynamics of underdamped particles driven by thermal fluctuations. In thermal equilibrium, its solution corresponds to the stationary velocity distribution by Maxwell, \emph{i.e.}, $P_{0}(\boldsymbol{v})=m[2\pi k_{B}T]^{-1} \exp\left\{-mv^{2}/2k_{B}T\right\}$. 

For the case $\Gamma(v)=\Gamma_{1} v,$ which decouples the dynamics of $v$ and $\theta$ as can be checked by simple inspection from \eref{speed} and \eref{direction}, we have straightforwardly that $P(v,\theta,t)=P(v,t)P(\theta,t).$ The stability analysis against velocity-alignment can be done along the lines of \cite{StrogatzPRL1992}.

For the case studied here, $\Gamma$ equal to a constant, the stability of the Maxwellian distribution $P_{0}(\boldsymbol{v})$ against velocity-alignment is proved below, by doing a stability analysis of the \emph{nonlinear} Fokker-Planck equation \eref{NonLinearKFP1}. We now introduce the \emph{ansatz} \cite{StrogatzJSP1991} for linear stability analysis:
\begin{equation}
P_{ans}(\boldsymbol{v},t)=P_{0}(\boldsymbol{v})+e^{\mu t}Q(\boldsymbol{v}).
\end{equation}
The normalization of $P_{ans}(\boldsymbol{v},t)$ implies that $Q$ must satisfy the condition 
\begin{equation}\label{Qcondition}
\int d^{2}\boldsymbol{v}\, Q(\boldsymbol{v})=0. 
\end{equation}
After substituting $P_{ans}$ in (\ref{NonLinearKFP1}), neglecting non-linear terms in $Q(\boldsymbol{v})$, and using the polar coordinates $v$ and $\theta$ for the velocity, we have in dimensionless variables (quantities with tilde),
\begin{equation}\label{StabAnal}
\fl\left\{\tilde{\mu}-\frac{1}{\tilde{v}}\frac{\partial}{\partial \tilde{v}}\tilde{v}^{2}-\frac{1}{2}\left(\frac{\partial^{2}}{\partial \tilde{v}^{2}}+\frac{1}{\tilde{v}}\frac{\partial}{\partial \tilde{v}}+\frac{1}{\tilde{v}^{2}}\frac{\partial^{2}}{\partial \theta^{2}}\right)\right\} \tilde{Q}(\tilde{v},\theta)=\frac{\widetilde{\Gamma}}{\tilde{v}} e^{-\tilde{v}^{2}}\, \frac{\Lambda}{2\pi} \cos\left(\psi-\theta\right).
\end{equation}
Since $\tilde{Q}(\tilde{v},\theta)$ is a 2$\pi$-periodic function in $\theta,$ we take its Fourier expansion
$\tilde{Q}(\tilde{v},\theta)=e^{-\tilde{v}^{2}}\sum_{n=-\infty}^{\infty}f_{n}(\tilde{v})e^{in\theta}.$ By using the variable $u=\tilde{v}^{2}$ and after substituting in (\ref{StabAnal}), multiplying by $\frac{1}{2\pi}e^{-in\theta}$ and integrating over $\theta$ from $0$ to $2\pi$ we get, 
\begin{eqnarray}\label{Equationfr}
\fl u\tilde{f}^{\prime\prime}_{n} +(1-u)\tilde{f}^{\prime}_{n}-\left(\frac{n^{2}}{4u}+\frac{\tilde{\mu}}{2}\right)\tilde{f}_{n}=-\frac{\widetilde{\Gamma}}{8\sqrt{u}}\times\nonumber\\
\qquad\qquad\qquad\left[\delta_{n,1}\int_{0}^{\infty}du^{\prime}\, e^{-u^{\prime}}\tilde{f}_{1}(u^{\prime})+\delta_{n,-1}\int_{0}^{\infty}du^{\prime}\, e^{-u^{\prime}}\tilde{f}_{-1}(u^{\prime})\right],
\end{eqnarray}
where we have used that the average of equation (\ref{ComplexOP}) implies
\begin{equation}
 \langle\Lambda\rangle e^{\pm i\langle\psi\rangle}=\pi\int_{0}^{\infty}du\, e^{-u}\tilde{f}_{\pm1}(u). 
\end{equation}

The condition \eref{Qcondition} involves only the $n=0$ Fourier mode, explicitly
\begin{equation}\label{Qcondition2}
 \int_{0}^{\infty}d\tilde{v}\, \tilde{v}e^{-\tilde{v}^{2}}f_{0}(\tilde{v})=0
\end{equation}
and the solution of equation (\ref{Equationfr}) for $n=0$ is given by
\begin{equation}
 f_{0}(\tilde{v})=M(\tilde{\mu}/2,1,\tilde{v}^{2}),
\end{equation}
$M(a,b,z)=\sum_{n=0}^\infty \frac {a^{(n)} z^n} {b^{(n)} n!}={}_1F_1(a;b;z)$ being the Kummer's function of the first kind and $(a)_{n}=a(a+1)\cdots(a+n-1)$ denote the Pochhammer symbol. 

After substituting $f_{0}$ in \eref{Qcondition2}, one gets $\sum_{n=0}^{\infty}\left(\tilde{\mu}/2\right)_{n}/n!=0,$ condition satisfied only for $\tilde{\mu}<0$. This corresponds to a necessary, but not sufficient, condition for the stability of the equilibrium distribution $P_{0}(\boldsymbol{v})$. 


\begin{figure}
\includegraphics[width=0.6\textwidth]{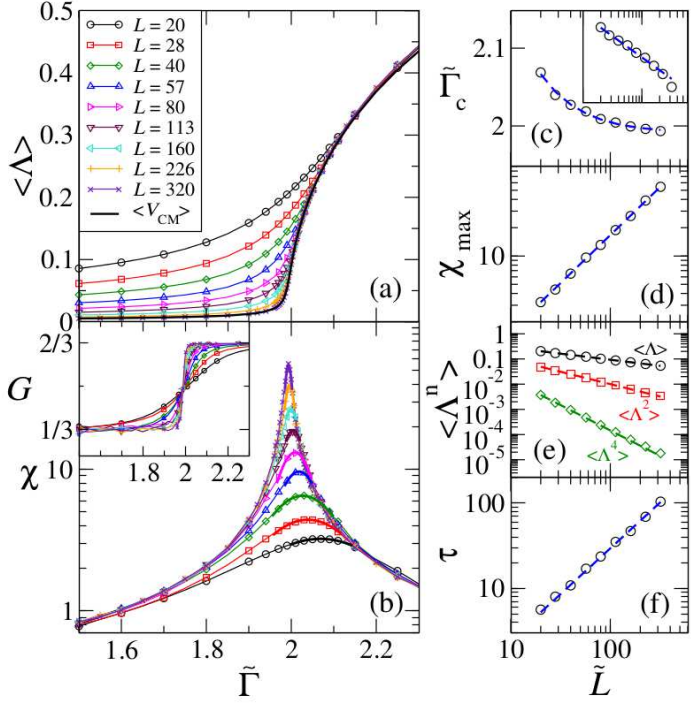}
\caption{(Color online) (a) Stationary order parameter $\langle \Lambda \rangle$ vs $\tilde{\Gamma}$ in the globally coupled regime $(R/L=1)$ for different system sizes. The solid lines in the curves with symbols are guides to the eye. The solid black line shows the stationary magnitude of the velocity of the center of mass, $\langle V_{\hbox{\tiny CM}}\rangle$, for $\tilde{L}=320$. Notice that $V_{\hbox{\tiny CM}}$ is a little smaller than $\langle \Lambda \rangle$ as $\tilde{\Gamma}$ increases. (b) Log-lin plot of the susceptibility $\chi$ vs $\tilde{\Gamma}$ for the systems shown in (a). The inset shows the corresponding Binder cumulant $G$ vs $\tilde{\Gamma}$; a universal crossing point could not be determined (see text). The log-log plots in (c) and (d) correspond to the location $\tilde{\Gamma}_c$ of the maximum of the susceptibility and the maximum of the susceptibility $\chi_{\hbox{\tiny max}}$ itself as a function of $\tilde{L}$, respectively. The clear circles were calculated from Gaussian fits (thick solid lines) to the crests of the the curves in (b). The dashed line in (c) corresponds to a power law fit from which the critical point $\tilde{\Gamma}_{c}^{\infty}=1.991(3)$ was determined. The inset in (c) shows the same data from which $\tilde{\Gamma}_{c}^{\infty}=1.99102$ has been subtracted, while the dashed line marks a power-law decay from which the critical exponent $\nu=0.94(4)$ was determined. The dashed line in (d) has a slope of $1.02(1)$. (e) Log-log plot of the moments $\langle \Lambda^n \rangle$ as a function of $\tilde{L}$ for $n=1,2,4$. The dashed lines correspond to power law fits that yield $\beta/\nu=0.491(5)$. (f) Log-log plot of the critical correlation time $\tau$ as a function of $\tilde{L}$. The dashed line has a slope of $1.07(2)$. In all cases $\tilde{\rho} = N/ \tilde{L}^2 =1$.}
\label{fig2}
\end{figure}

In order to estimate the critical exponents associated with the phase transition as well as the critical value of the coupling constant, we have performed a standard finite-size-scaling analysis \cite{GinelliPRL2010}. For our numerical simulations, we have discretized equations \eref{pos_eq} and \eref{vel_eq} as well as (\ref{Aligning}). In its dimensionless form and for the global coupling case, the control parameter is $\tilde{\Gamma}$ for any given $\tilde{L}$ and $\tilde{\rho}$. Without any loss of generality, we choose $\tilde{\rho}=1$. All numerical results presented here were obtained integrating these equations with a modified version of the velocity-Verlet algorithm \cite{GrootJCP1997} with an integration time-step $\Delta t = 0.01$. The results are shown in figure \ref{fig2}, where the stationary values of the order parameter, $\langle \Lambda\rangle$, the susceptibility \hbox{$\chi=\tilde{L}^d(\langle \Lambda^2\rangle-\langle \Lambda\rangle^2)$}, and the Binder cumulant \hbox{$G=1-\langle \Lambda^4\rangle/(3\langle \Lambda^2\rangle^2)$} are plotted, vs $\tilde{\Gamma}$, in figures \ref{fig2}(a), \ref{fig2}(b) and the inset of figure \ref{fig2}(b), respectively, for different system sizes. We were not able to determine uniquely the critical point from the crossing of the Binder cumulant curves. Instead, its value in the thermodynamic limit, $\tilde{\Gamma}_c^{\infty}$, and the critical exponent $\nu$ were determined from the scaling with $\tilde{L}$ of the location $\tilde{\Gamma}_c(\tilde{L})$ of the maximum of the susceptibility and the maximum of the suscepti\-bi\-li\-ty itself, $\chi_{\hbox{\scriptsize max}}(\tilde{L}) \propto (\tilde{\Gamma}_c(\tilde{L})-\tilde{\Gamma}_c^{\infty})^{-\gamma} \propto \tilde{L}^{\gamma/\nu}$, 
that lead to $\tilde{\Gamma}_c(\tilde{L}) = \tilde{\Gamma}_c^{\infty} + a\tilde{L}^{-1/\nu}$ (here, $\gamma$ refers to the critical exponent). From these expressions, we estimated $\tilde{\Gamma}_{c}^{\infty}=1.991(3)$, that is compatible with the one obtained for the Kuramoto model $\tilde{\Gamma}_{c}^{\infty}=2$, $\nu=0.94(4)$ and $\gamma/\nu=1.02(1)$ [see figures \ref{fig2}(c) and \ref{fig2}(d)].

Going further with the analysis, the scaling of the moments of the order pa\-ra\-me\-ter $\langle \Lambda^n\rangle \propto \tilde{L}^{-n(\beta/\nu)}$, shown in figure \ref{fig2}(e) for $n=1,2,4$, yields $\beta/\nu=0.491(5)$ from which $\beta=0.46(2)$. The so-called hyperscaling relation \hbox{$2\beta/\nu + \gamma/\nu = d$} is nicely confirmed with an estimate $d=2.00(1)$. We also computed the correlation time $\tau(\tilde{\Gamma},\tilde{L})$ from the exponential decay of the autocorrelation function of the order parameter. From the scaling relation $\tau( \tilde{\Gamma}_c^{\infty},\tilde{L}) \propto \tilde{L}^z$, shown in figure \ref{fig2}(f), we estimated the dynamical exponent $z=1.07(2)$. It is worth to mention that we accumulated our stationary values and distributions by integrating the equations of motion in the stationary state at least $10^3 \tau$.

Clearly, the critical exponents obtained are in agreement with the mean-field values known for the Kuramoto model, for instance, $\beta=\frac{1}{2}$. We may attribute this agreement to the fact that the dynamics of the velocity direction, $\theta,$ is weakly coupled to the dynamics of speed, driven by thermal fluctuations. Notice that the ratios of the critical exponents $\beta/\nu$ and $\gamma/\nu$ are very close to simple integer ratios as expected for mean field models.

On a side note, in order to determine the nature of the phase transition displayed by our model, one has to look at the behavior of the Binder cumulant.  As expected, the Binder cumulant curves, shown in the inset of figure \ref{fig2}(b), approch the values $\frac{1}{3}$ on the disordered (left) side and $\frac{2}{3}$ on the ordered (right) one. Moreover, their continuous behavior that never acquires negative values is a signature that we are dealing here with a second order (continuous) phase transition \cite{BinderRPP1997}.

\section{\label{conclusions} Conclusions}

In summary, our results show that non-Hamiltonian interactions that do not preserve momentum (such as the alignment interaction typically used to model flocking behavior), drive the system from an equilibrium to an out-of-equilibrium phase. This contrast with previous results where these transitions occur in systems either in equilibrium or out-of-equilibrium throughout the transition. Thus, our model is sui\-ta\-ble for studying the passage from equilibrium to non-equilibrium states, in particular, for the study of the passage from maximum entropy (equilibrium) states to stationary phases where entropy is produced. It is worth noting that a natural generalization of this study would include active fluctuations which refer to non-equilibrium fluctuations, along the direction of motion (speed noise) and perpendicular to it (angular noise), respectively. We are currently pursuing this line of investigation.

Additionally, we have also shown a clear relation between synchronization and collective motion, in particular with the Kuramoto model for synchronization. Even though collective motion is expected for the globally coupled case studied here (the local case is reported elsewhere), by not neglecting the coupling between synchronizing behavior and the motion of the particles, we have also shown that the transition exists for finite values of the control parameter and in contrast to previous results for flocking, where this transition occurs only at the maxi\-mum noise intensity (i.e., in the infinite temperature limit) but always far from equilibrium.

Beyond the obvious implications in the study of synchronization and flocking phenomena, we believe our results are relevant in the general context of phase transitions whether in equi\-li\-brium or out-of-equilibrium. 


\ack 
The authors gratefully acknowledge the computing time granted on the super\-com\-pu\-ters MIZTLI (DGTIC-UNAM), THUBAT-KAAL (CNS-IPICyT) and, through the project ``Cosmolog\'{\i}a y as\-tro\-f\'{\i}\-si\-ca relativista: objetos compactos y materia obs\-cu\-ra'', on \hbox{XIUHCOATL} (\hbox{CINVESTAV}). F.J.S.\ acknowledges support from the grant PAPIIT-IN113114. V.D.\ acknowledges support from the grant PROMEP/103.5/10/7296 and from CONACyT.

\end{document}